\documentclass[prd,12pt,nofootinbib,superscriptaddress]{revtex4-2}
\usepackage{bm}
\usepackage{amsmath}
\usepackage{amssymb}
\usepackage{graphicx}
\usepackage{float}
\usepackage{subfigure}
\usepackage{hyperref}
\hypersetup{
	colorlinks=true,
	linkcolor=red,
	citecolor=blue,
}

\begin{document}
\title{Rotating galactic black holes}
\author{Chao Zhang}
\email{zhangchao666@sjtu.edu.cn}
\affiliation{School of Physics and Astronomy, Shanghai Jiao Tong University, 800 Dongchuan Rd, Shanghai 200240, China}

\author{Guoyang Fu}
\email{fuguoyangedu@sjtu.edu.cn}
\affiliation{School of Physics and Astronomy, Shanghai Jiao Tong University, 800 Dongchuan Rd, Shanghai 200240, China}

\author{Chunyu Zhang}
\email{Corresponding author. chunyuzhang@itp.ac.cn}
\affiliation{CAS Key Laboratory of Theoretical Physics, Institute of Theoretical Physics, Chinese Academy of Sciences, Beijing 100190, China}

\begin{abstract}
The galactic black hole is a supermassive black hole located at the center of a galaxy surrounded by a dark matter halo.
For the first time, we establish a generic, fully-relativistic formalism to calculate solutions of Einstein's gravity minimally coupled to an anisotropic fluid modeled by the Einstein cluster in axisymmetric, non-vacuum spacetimes, which are extensions of spherically-symmetric cases.
These asymptotically flat spacetimes with regular horizons can describe the geometry of galaxies harboring supermassive black holes and are useful to constrain the environment surrounding astrophysical black holes.
Our findings provide a solid groundwork for future studies on the shadows, quasi-normal modes, and other phenomena associated with rotating galactic black holes.
\end{abstract}

\maketitle

\section{Introduction}
There is overwhelming evidence supporting the existence of dark matter (DM) by astrophysical and cosmological observations over various ranges of scales, such as the rotation curves \cite{Rubin:1970zza, Begeman:1991iy}, bullet clusters \cite{Zwicky:1933gu, Clowe:2006eq}, large-scale structures \cite{Dietrich:2012mp}, and cosmic microwave background anisotropics \cite{Planck:2018vyg}.
A supermassive galactic black hole (GBH) of mass from $10^6~M_{\odot}$ to $10^{10}~M_{\odot}$ usually exists at the center of the galaxy, surrounded by a DM halo that is substantially larger than the visible galaxy \cite{Kormendy:2013dxa, Harris:2015vxa}.
As a matter of fact, the Sagittarius $A^*$ black hole (BH) at the centre of Milky Way galaxy \cite{EventHorizonTelescope:2022wkp} and $\rm M87^*$ BH at the centre of Messier $87$ galaxy \cite{EventHorizonTelescope:2019dse} have been already observed by the Event Horizon Telescope.

Diverse space-time models are employed to match an isolated BH with a certain density distribution of DM halo.
The supermassive black hole (SMBH) at the center of a galaxy, in relation to a specific distribution of DM halo, was initially investigated through the application of Newtonian gravity \cite{Gondolo:1999ef}.
In 1939 Einstein presented a stationary system of a thick spherical shell composed of many equal gravitating masses, each moving in a circular geodesic orbit in the field of all the other \cite{Einstein:1939ms}.
Such a system has spherical symmetry and is referred to in the literature as an "Einstein cluster".
Boehmer and Harko \cite{Boehmer:2007az} and Lake \cite{Lake:2006pp}, as well as other related studies \cite{GLComer_1993,Szybka:2018hoe,Cardoso:2021wlq,Figueiredo:2023gas,Speeney:2024mas,Cardoso:2022whc,Jusufi:2022jxu,Shen:2023erj,Geralico:2012jt} used Einstein cluster to explain DM in a fully-relativistic formalism.
Cardoso et al \cite{Cardoso:2021wlq, Cardoso:2022whc} worked out the spacetime geometry with a spherically symmetric, static, non-vacuum BH generated by the DM distribution with the Hernquist density profile \cite{1990ApJ356359H}.
This solution quickly garnered significant interest, with its diverse properties and applications being extensively explored.
These include Love numbers, quasinormal modes, tails, grey-body factors \cite{Konoplya:2021ube,Liu:2022lrg, Destounis:2022obl}, and shadows \cite{Xavier:2023exm}, as well as the detectability of gravitational waves (GWs) emitted by extreme mass ratio inspirals by observatories such as LISA \cite{Danzmann:1997hm,LISA:2017pwj}, TianQin \cite{Luo:2015ght}, and Taiji \cite{Hu:2017mde}.
Recently, three analytic energy densities of the halo satisfying all the energy conditions, weak, strong, and dominant \cite{Hawking:1973uf} have been constructed to represent a SMBH located at the center of a galaxy surrounded by a DM halo \cite{Shen:2023erj}.
The distinction between these models is primarily found in the slopes of the density profiles within the spike and distant regions.
However, the generalization of spherical GBH to rotational GBH has not been carried out even for the Hernquist density profile due to the complexity of the problem.

Applications of rotating solutions to astrophysics and theories of gravity are of great importance.
The Newman-Janis algorithm \cite{Newman:1965tw} was first devised to generate exterior rotating solutions from the spherical symmetric space-time metric.
Utilizing the Newman-Janis method, the analytical form of BH space-time metric in DM halo for the stationary situation (spherical symmetric and rotational) was obtained in \cite{Xu:2018wow, Hou:2018avu}.
Regrettably, the above DM model is the phenomenological perfect fluid DM model \cite{Kiselev:2002dx, Rahaman:2010xs}, wherein the DM is characterized as a perfect fluid.
An approximate solution for the metric exists to describe the galactic halo scenario, achieved by selecting metric functions that fulfill specific conditions $g_{tt}=-g_{rr}^{-1}$ \cite{Li:2012zx}.
The Newman-Janis algorithm cannot be applied to derive the rotating GBH metric surrounded by the anisotropic DM fluid modeled by the Einstein cluster for the reason that the Newman-Janis algorithm only works under the conditions $g_{tt}=-g_{rr}^{-1}$ \cite{Azreg-Ainou:2014aqa, Viaggiu:2006mx, HJ1982}.
Unsurprisingly, the generalization of spherical symmetric GBH to rotating GBH surrounded by Einstein cluster typically increases the complexity of the field equations to such a degree that the analytical solution becomes incredibly difficult and sometimes even seemingly impossible.
We introduce a numerical framework capable of generating exact solutions that depict stationary, axisymmetric GBH within a DM halo.
The code is based on a relaxed Newton-Raphson method to solve the discretized field equations with Newton's polynomial finite difference scheme developed by Sullivan et al \cite{Sullivan:2019vyi,Sullivan:2020zpf}.

The paper is organized as follows.
In Sec. \ref{sec2}, we briefly review the Einstein clusters constructed by Einstein and outline the numerical algorithm as applied to rotating GBH.
In Sec. \ref{sec3}, we validate the algorithm by considering Kerr BHs and spherically symmetric GBH.
In Sec. \ref{sec4}, we apply the code to rotating GBH and derive the results.
Sec. \ref{sec5} is devoted to conclusions and discussions.
In this paper, we set $G=c=1$.

\section{Formalism}\label{sec2}
The surrounding anisotropic matter, according to the Einstein cluster, can be expressed in terms of the average stress tensor \cite{Einstein:1939ms,Boehmer:2007az,GLComer_1993}
\begin{equation}
T^{\mu\nu}=\frac{n}{m_p}\left\langle P^\mu P^\nu\right\rangle,
\end{equation}
where $n$ is the number density of particles with mass $m_p$ and $P^\mu$ the four-momentum satisfying the geodesic equations, and the averaging $\langle ...\rangle$ is performed over all possible trajectories going through the generic spatial point where the stress-energy tensor is evaluated, i.e., overall directions and phase.
Let us start with an axisymmetric and stationary metric ansatz in isotropic coordinates
\begin{equation}\label{metric}
ds^2 = - f (\rho, \theta) d t^2 + \frac{m (\rho, \theta)}{f (\rho, \theta)} \left( d \rho^2 + \rho^2 d \theta^2 \right) 
+ \frac{l (\rho, \theta)}{f (\rho, \theta)} \rho^2 \sin^2\theta \left( d \phi - \frac{\omega (\rho, \theta)}{\rho} d t \right)^2,
\end{equation}
where $\rho$ is the isotropic radial coordinate.
The event horizon of the static BH solutions is characterized by $g_{tt}=-f=0$, $g_{\rho\rho}$ is finite at the horizon in isotropic coordinates.
We stipulate that the horizon of the BH solutions is located at a surface where $\rho = \rho_H$ is constant.
This ansatz for the event horizon is justified retrospectively, as it yields consistent solutions with a regular event horizon \cite{Kleihaus:1997ws}.
Due to symmetry properties, we employ an orthonormal basis 
\begin{align}
&e^t_\mu=\left(\sqrt{f},0,0,0\right),\qquad &e^\rho_\mu=\left(0,\sqrt{\frac{m}{f}},0,0\right),\\
&e^\theta_\mu=\left(0,0,\rho\sqrt{\frac{m}{f}},0\right),\qquad &e^\phi_\mu=\left(-\omega\sin\theta\sqrt{\frac{l}{f}},0,0,\rho\sin\theta \sqrt{\frac{l}{f}}\right),
\end{align}
where $e_t^\mu$ is the 4-velocity vector of the fluid.
The general imperfect energy tensor can be represented by
\begin{equation}
    T_{\mu\nu}=\epsilon \, e^t_\mu\, e^t_\nu+p_\rho \,e^\rho_\mu\, e^\rho_\nu+ p_\theta\, e^\theta_\mu\, e^\theta_\nu+p_\phi \,e^\phi_\mu \,e^\phi_\nu\,
\end{equation}
where $\epsilon$ is the matter density and $(p_\rho,p_\theta,p_\phi)$ are the components of the pressure.
According to the Einstein cluster, the test particles travel along the thin shell with no pressure in the radial direction.
After performing the averaging, the density $\epsilon$ and pressures $(p_\rho,p_\theta,p_\phi)$ will depend only on the radius $\rho$.
Especially, the radial pressure will vanish, $p_\rho=0$.
In the case of axisymmetry, the tangential pressure $p_\theta$ may not equal $p_\phi$, where $p_\theta=p_\phi$ holds for the spherical case.
The final Einstein metric field equation yields
\begin{equation}
H_{\mu\nu}=G_{\mu\nu}-T_{\mu\nu}=0.
\end{equation}
To simplify the partial differential equations, following \cite{Kleihaus:2015aje,Sullivan:2020zpf}, we use linear combinations of the tensor $H_{\mu\nu}$ to diagonalize the equations with respect to the operator $\hat{\mathcal{O}}=\frac{\partial^2}{\partial \rho^2}+\frac{1}{\rho^2}\frac{\partial^2}{\partial \theta^2}$,
\begin{equation}
\begin{split}
\frac{m}{f}\left(H^\mu_{~\mu}-2H^t_{~t}-\frac{2\omega}{\rho}H^t_{~\phi}\right)&=\frac{1}{f}\hat{\mathcal{O}}f+\dots,\\
2\frac{m}{f}\left(H^\phi_{~\phi}-\frac{\omega}{\rho}H^t_{~\phi}\right)&=\frac{1}{m}\hat{\mathcal{O}}m+\dots,\\
2\frac{m}{f}\left(H^\rho_{~\rho}+H^\theta_{~\theta}\right)&=\frac{1}{l}\hat{\mathcal{O}}l+\dots,\\
2\frac{fm}{l\sin^2\theta}\left(-\frac{1}{\rho}H^t_{~\phi}\right)&=\hat{\mathcal{O}}\omega+\dots,\\
\end{split}
\end{equation}
The conservation of matter equation $\nabla_\mu T^{\mu\nu}=0$ yields two constraint
\begin{equation}
\begin{split}
p_\theta&=p_\theta(\epsilon,f,m,l,\omega),\\
p_\phi&=p_\phi(\epsilon,f,m,l,\omega),
\end{split}
\end{equation}
where $(p_\theta,p_\phi)$ can be expressed in terms of functions $(\epsilon,f,m,l,\omega)$ and their derivative.
For a Kerr BH with mass $M_0$ and dimensionless spin $\chi_0$, the isotropic coordinate $\rho$ is related to the Boyer-Lindquist radial coordinate by \cite{Sullivan:2020zpf}
\begin{equation}\label{BLI}
r=\rho+M_0+\frac{M_0^2-M_0^2\chi_0^2}{4\rho}.
\end{equation}
It is convenient to replace the dimensionless spin parameter $\chi_0$ with the event horizon radius $\rho_H$ using the relation \cite{Sullivan:2020zpf}
\begin{equation}
\chi_0=\frac{\sqrt{M_0^2-4\rho_H^2}}{M_0}.    
\end{equation}
To obtain asymptotically flat solutions with a regular event horizon and with the proper symmetries, we must impose the appropriate boundary conditions, the boundaries being the horizon and radial infinity, the z-axis, and the $\rho$-axis because of parity reflection symmetry.
Requiring the horizon to be regular, we obtain the boundary conditions at the horizon $\rho=\rho_H$.
The metric functions must satisfy
\begin{equation}
f|_{\rho = \rho_H}=m|_{\rho = \rho_H}=l|_{\rho = \rho_H}=0, \qquad \omega|_{\rho=\rho_H}=\omega_{\rm Kerr}|_{\rho=\rho_H}=\frac{\rho_H\sqrt{M_0^2-4\rho_H^2}}{2M_0(M_0+2\rho_H)}.
\end{equation}
At infinity, we require the boundary conditions
\begin{equation}\label{BC2}
f|_{\rho = \infty}=m|_{\rho = \infty}=l|_{\rho = \infty}=1,\qquad \omega|_{\rho=\infty}=0.
\end{equation}
The symmetries determine the boundary conditions along the $\rho$-axis and the $z$-axis
\begin{equation}\label{BC3}
\begin{split}
   \partial_{\theta}f|_{\theta=0}= &\partial_{\theta}m|_{\theta=0}=\partial_{\theta}l|_{\theta=0}=\partial_{\theta}\omega|_{\theta=0}=0,  \\
\partial_{\theta}f|_{\theta=\pi/2}= &\partial_{\theta}m|_{\theta=\pi/2}=\partial_{\theta}l|_{\theta=\pi/2}=\partial_{\theta}\omega|_{\theta=\pi/2}=0  .
\end{split}
\end{equation}
To prepare our field equations for numerical integration, we make an additional substitution \cite{Sullivan:2020zpf}
\begin{equation}
    x=1-\frac{\rho_H}{\rho}.
\end{equation}
This substitution changes our domain of integration from $\rho\in [\rho_H,\infty)$ to the finite domain $x\in [0,1]$.
Another substitution is necessary to eliminate a numerical divergence on the event horizon.
We replace the metric functions with corresponding barred functions defined by
\begin{equation}\label{funcs}
\begin{split}
f&=x^2\bar{f},\\
m&=x^2\bar{m},\\
l&=x^2\bar{l},\\
\omega&=\bar{\omega}.
\end{split}
\end{equation}
This substitution leaves the boundary conditions as $x\to 1$ unchanged.
At the horizon, the boundary conditions are obtained from examining an expansion of the metric functions around $x=0$ and become
\begin{equation}\label{BCfinal}
\begin{split}
  \left( \bar{f}-\frac{\partial \bar{f}}{\partial x} \right)|_{x=0}&=0 ,\\
\left( \bar{m}+\frac{\partial \bar{m}}{\partial x} \right)|_{x=0}&=0 ,\\
\left( \bar{l}-\frac{\partial \bar{l}}{\partial x} \right)|_{x=0}&=0 ,\\
   \bar{\omega}|_{x=0}  & =\omega_H.
\end{split}
\end{equation}
For the DM density of the Hernquist profile given in the Boyer-Lindquist coordinate \cite{Cardoso:2021wlq},
we can use the coordinate transformation in Eq. \eqref{BLI} to approximately give the DM density in the isotropic coordinate
\begin{equation}
\begin{split}
\epsilon(x)=&-\frac{M_{\rm DM} \rho_H (x-1)^4 x^2 }{2 \pi  \left(M_0 (x-1)-\rho_H \left(x^2-2 x+2\right)\right)^2 } \times  \\
&\qquad \qquad\qquad \frac{(L_{\rm DM}+2 M_0)}{ \left(L_{\rm DM} (x-1)+M_0 (x-1)-\rho_H \left(x^2-2 x+2\right)\right)^3}\,,
\end{split}
\end{equation}
where $M_{\rm DM}$ is the total mass of the halo and $L_{\rm DM}$ is a typical lengthscale.

\section{Validation}\label{sec3}
We discretize our differential operators using their Newton polynomial representation of order $16$ on a 2-dimensional grid $61\times 31$ points and initialize our solver with the initial guess \cite{Sullivan:2020zpf}.
The four input parameters that we must specify are the horizon radius where we choose $\rho_H$ and the dimensionless spin parameter $\chi_0$.
These two input parameters are chosen to determine the property of the BH.
The other two parameters are $M_{\rm DM}$ and $L_{\rm DM}$, which determine the DM distribution around the BH.
For all computations in this paper, we set $\rho_H=1$.
The horizon angular velocity is chosen to coincide with that of a Kerr BH of the dimensionless spin parameter.
Our specified tolerance is chosen to be less than $10^{-5}$.

\begin{figure}
    \centering
    \includegraphics[width=0.96\columnwidth]{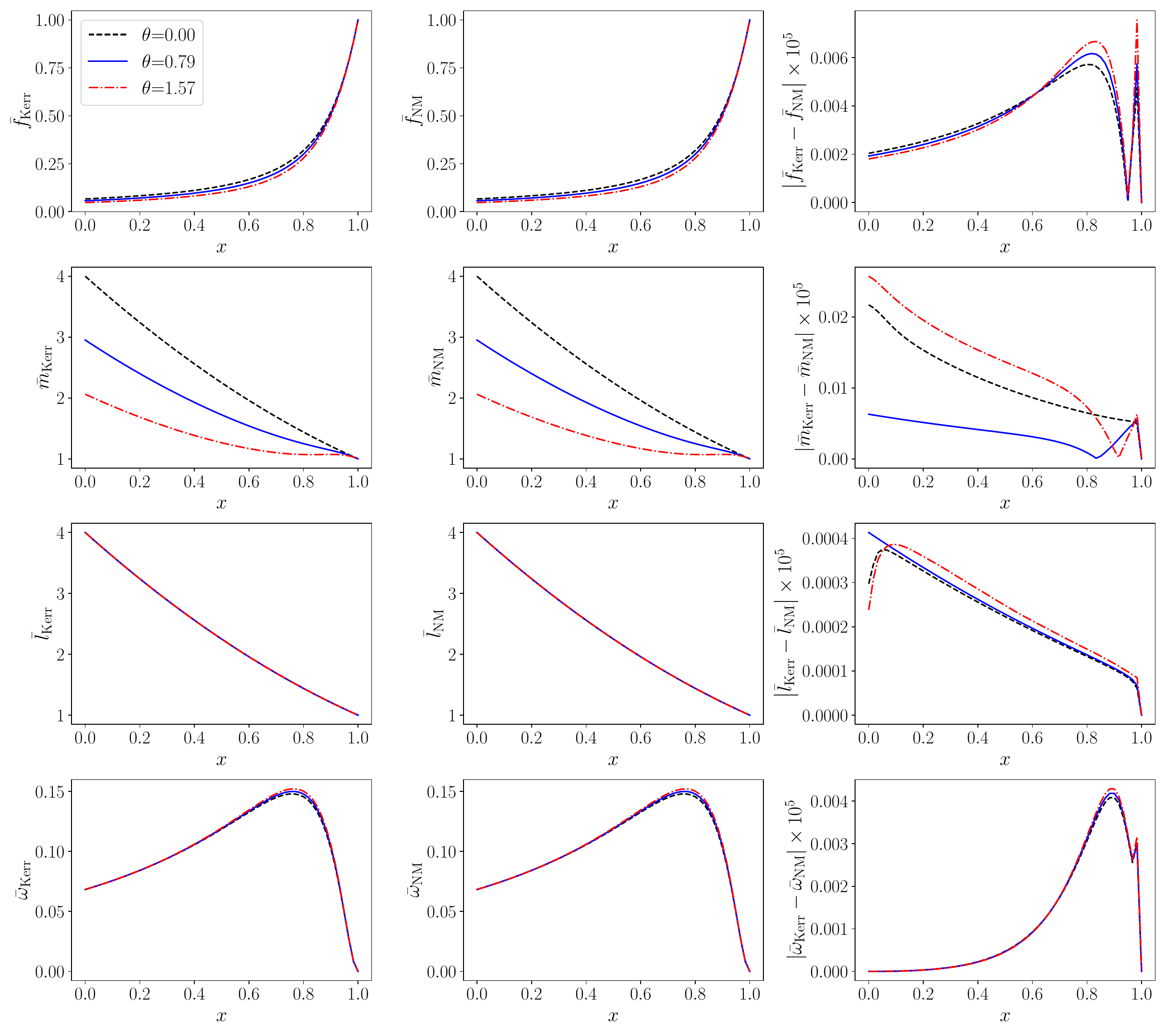}
    \caption{The numerical value, the analytical value, and the absolute error for each metric element to the Kerr solution for three selected angles.
     The term "Kerr" means the analytical results of each metric element while the term "NM" means the result from our numerical code.
    Here we show the metric components and the absolute error $\bar{f}$ (first row), $\bar{m}$ (second row), $\bar{l}$ (third row), and $\bar{\omega}$ (fourth row) and for three angles $\theta=0,0.79,1.57$ denoted by dashed, solid, and dotted lines respectively.
    The BH parameters are chosen $(\rho_H=1,\chi_0=0.9)$ and the DM parameter $(M_{\rm DM}=0, L_{\rm DM}=10)$.
    }
    \label{KERRvsNM}
\end{figure}

To validate our code, we choose the BH parameter $(\rho_H=1,\chi_0=0.9)$ and DM parameter $(M_{\rm DM}=0, L_{\rm DM}=10)$.
We find that our numerical infrastructure converges to the desired solution below our specified tolerance of $10^{-5}$.
The numerical result for the metric elements $(\bar{f},\bar{m},\bar{l},\bar{\omega})$ can be seen in Fig. \ref{KERRvsNM}.
For the Kerr limit, the metric elements $(\bar{f},\bar{m},\bar{\omega})$ possess a complicated shape with a nontrivial dependence on the angular coordinate $\theta$ while the metric element $\bar{l}$ is only dependent on the radial variable.
For given three angles $\theta=0,0.79,1.57$, the values of the metric elements $(\bar{f},\bar{m})$ are larger at the z-axis than the equatorial plane while the value of the metric element $\bar{\omega}$ is opposite.
It's obvious that the frame-dragging term $\bar{\omega}$ is larger at the equatorial plane rather than the z-axis.
The Kerr metric is given usually in Boyer-Lindquist coordinates, we rewrite it by employing the metric ansatz \eqref{metric} in Appendix \ref{KERRICS}.
We can compare each metric element value of the rotating GBH without DM with the analytical Kerr solutions to validate our code.
The numerical value, the analytical value, and their absolute error between each metric function are shown in Fig. \ref{KERRvsNM}.
It's also proved from the analytical form that the metric element $\bar{l}$ is only dependent on the radial variable and the metric elements $(\bar{f},\bar{m},\bar{\omega})$ have a dependence on the angular coordinate $\theta$.
This figure validates our numerical code to construct stationary and axisymmetric BH solutions without DM.

Another important limit is found by taking the spherically symmetric limit.
The solutions in this case are spherically symmetric and have been discussed in \cite{GLComer_1993,Szybka:2018hoe,Cardoso:2021wlq,Figueiredo:2023gas,Speeney:2024mas,Cardoso:2022whc,Jusufi:2022jxu,Shen:2023erj,Geralico:2012jt}.
For the spherically symmetric case, the corresponding ansatz in isotropic coordinates reads
\begin{equation}
ds^2=-a(\rho) dt^2+b(\rho) \left( d \rho^2 + \rho^2 d \theta^2+ \rho^2\sin^2\theta d \phi^2\right),
\end{equation}
and the anisotropic stress tensor is characterized by \cite{Cardoso:2021wlq}
\begin{equation}
T^{\mu}_{~\nu}={\rm diag}(-\epsilon,0, p_t,p_t), 
\end{equation}
where $\epsilon$ is the energy density of the fluid, $p_t$ is the tangential pressure.
We can follow the method presented by Cardoso et al \cite{Cardoso:2021wlq} to give the metric of a GBH surrounded by DM.
In order to compare the result from Cardoso et al's method and our numerical code, we make the substitution
\begin{equation}
f_{\rm SSC}(\rho)=a(\rho),\qquad m_{\rm SSC}(\rho)=a(\rho)b(\rho).
\end{equation}
The details of calculating metric elements $a(\rho)$ and $b(\rho)$ can be seen in Appendix \ref{SSCGBH}.
\begin{figure}
    \centering
    \includegraphics[width=0.9\columnwidth]{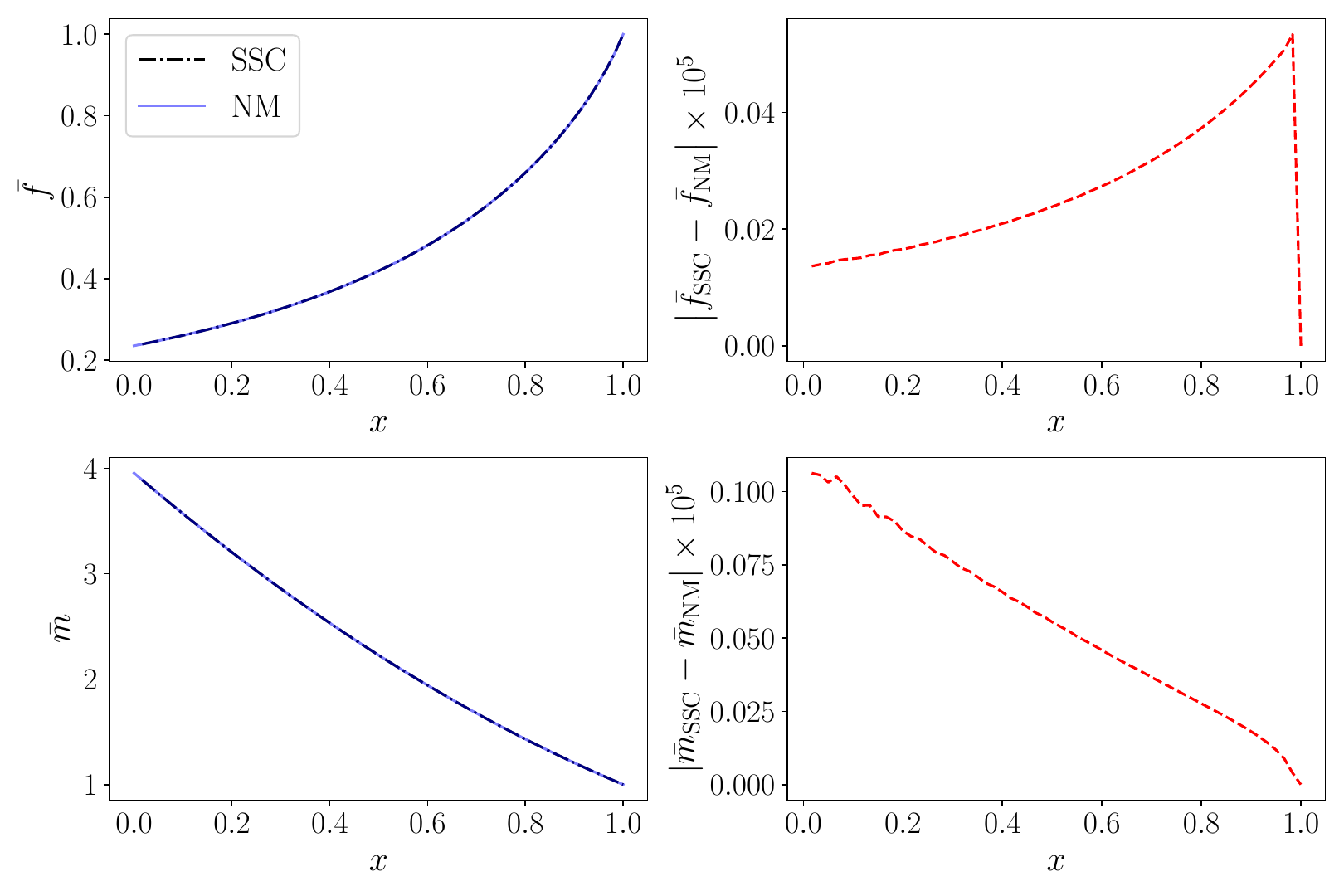}
    \caption{The value of each metric element from different methods and their absolute error. The term "SSC" means the results of each metric element from Cardoso's method while the term "NM" means the result from our numerical code.
    The BH parameters are chosen $(\rho_H=1,\chi_0=0)$ and the DM parameter $(M_{\rm DM}=10, L_{\rm DM}=10)$.}
    \label{SSCvsNM}
\end{figure}
To validate our code, we choose the BH parameter $(\rho_H=1,\chi_0=0)$ and DM parameter $(M_{\rm DM}=10, L_{\rm DM}=10)$.
The metric element for $\bar{f}$ and $\bar{m}$ can be seen in Fig. \ref{SSCvsNM}.
We can see that the metric elements calculated from Cardoso's method are the same as the result from our numerical code.
The absolute error from two different methods can satisfy our specified tolerance to be less than $10^{-5}$.
Figure \ref{SSCvsNM} validates our numerical code to construct stationary and spherical BH solutions with a given DM halo density.
Based on these numerical solutions, we can confidently use our numerical code to construct the rotating GBH within the specified tolerance.

\section{Rotating galactic black holes}\label{sec4}
\begin{figure}
    \centering
    \includegraphics[width=0.95\columnwidth]{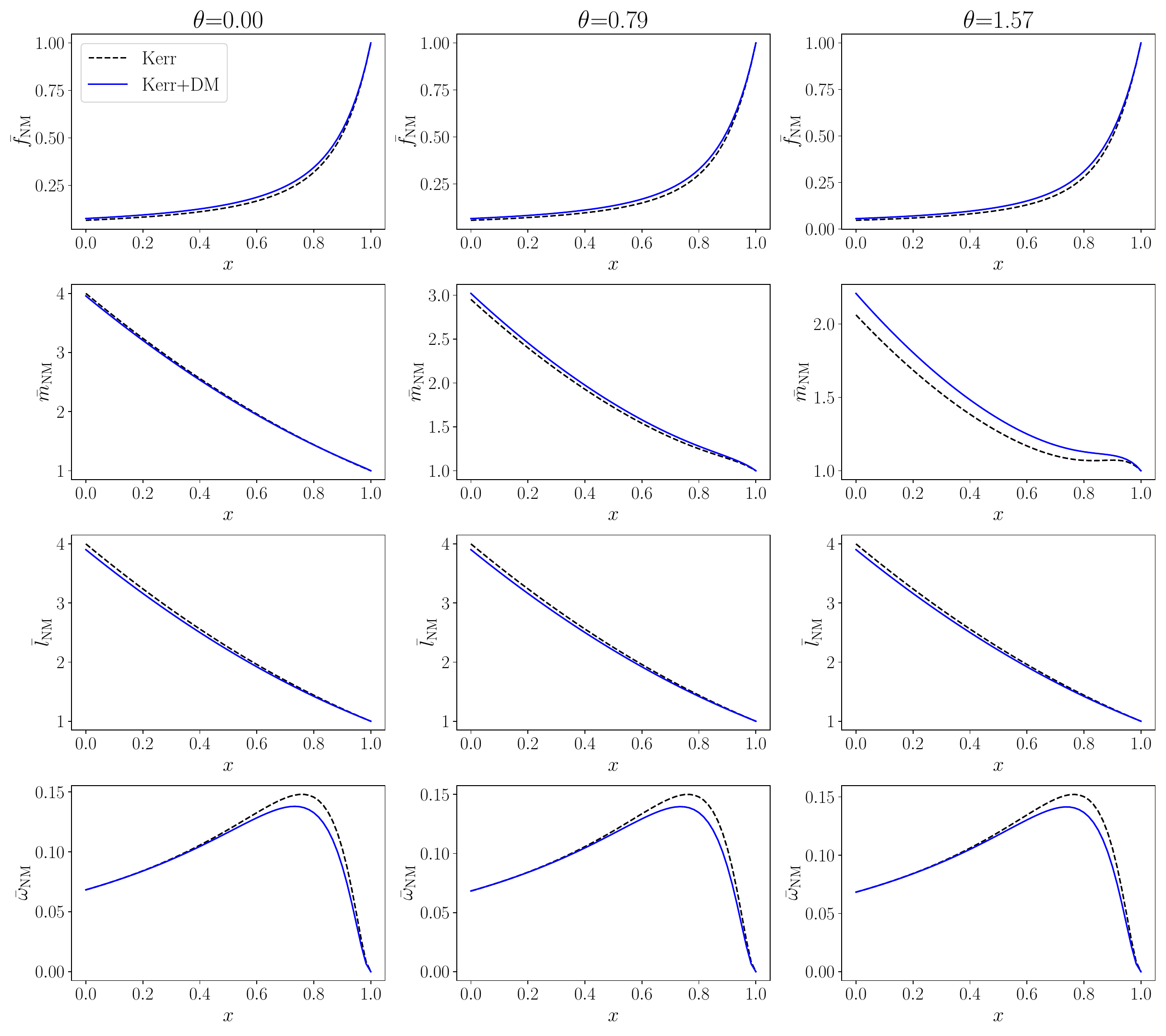}
    \caption{The value for each metric element for three selected angles.
     The term "Kerr" means the analytical results of each metric element without DM while the term "Kerr+DM" means the numerical result with DM from our numerical code.
    Here we show the metric components $\bar{f}$ (first row), $\bar{m}$ (second row), $\bar{l}$ (third row), and $\bar{\omega}$ (fourth row) and for three angles $\theta=0,0.79,1.57$.
    The BH parameters are chosen $(\rho_H=1,\chi_0=0.9)$ and the DM parameter $(M_{\rm DM}=10, L_{\rm DM}=10)$.}
    \label{DMKERR}
\end{figure}

\begin{figure}
    \centering
    \includegraphics[width=0.95\columnwidth]{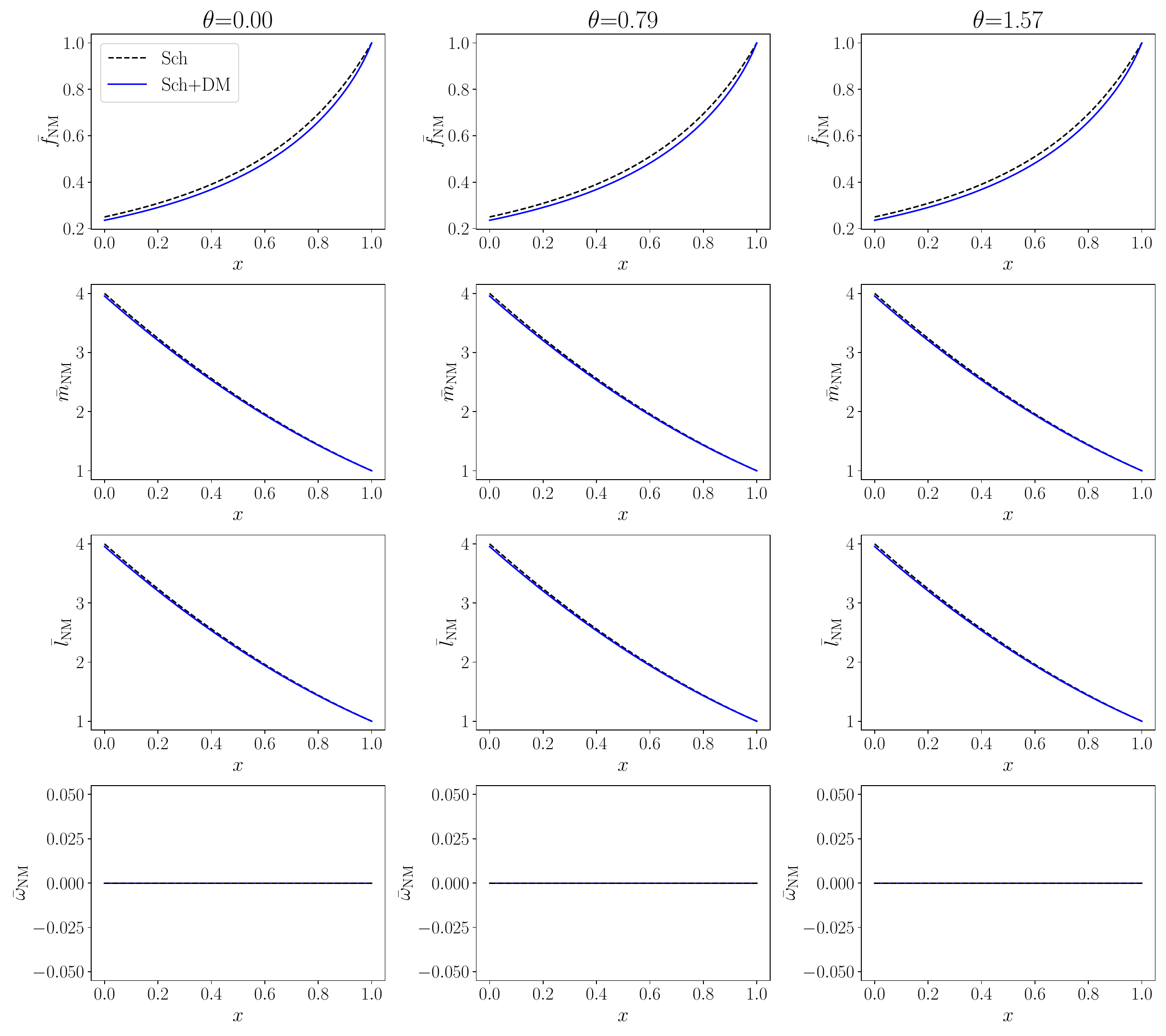}
    \caption{The value for each metric element for three selected angles.
     The term "Sch" means the analytical results of each metric element without DM while the term "Sch+DM" means the numerical result with DM from our numerical code.
    Here we show the metric components $\bar{f}$ (first row), $\bar{m}$ (second row), $\bar{l}$ (third row), and $\bar{\omega}$ (fourth row) and for three angles $\theta=0,0.79,1.57$.
    The BH parameters are chosen $(\rho_H=1,\chi_0=0)$ and the DM parameter $(M_{\rm DM}=10, L_{\rm DM}=10)$.}
    \label{SchvsDMSch}
\end{figure}

In this section, we numerically solve the Einstein equations with DM halo for the static axially symmetric BH solutions. 
We discretize our differential operators using their Newton polynomial representation of order $16$ on a 2-dimensional grid $61\times 31$ points.
Our specified tolerance is chosen to be less than $10^{-5}$.
We chose the four input parameters: the BH parameter chosen $(\rho_H=1,\chi_0=0.9)$ and DM parameter $(M_{\rm DM}=10, L_{\rm DM}=10)$.
The metric elements $(\bar{f},\bar{m},\bar{l},\bar{\omega})$ for the rotating GBH can be seen in Fig. \ref{DMKERR}.
Compared with the Kerr metric, the metric value of rotating GBH is smaller for the functions $(\bar{l},\bar{\omega})$ and larger for the function $\bar{f}$ no matter what the angles are.
For the function $\bar{m}$, the difference between rotating GBH and the Kerr metric is highly dependent on the angle.
At the z-axis, the value of $\bar{m}$ for GBH is smaller than the value for the Kerr case.
However, the value of $\bar{m}$ for GBH is larger than the value for the Kerr case at the equatorial plane.
For the frame-dragging term $\bar{\omega}$, the difference between rotating GBH and the Kerr BH will increase to the highest and then decrease to zero as the radius from the horizon to infinity.
It is obvious that the frame-dragging term $\bar{\omega}$ will tend to zero at infinity no matter whether DM exists or not.
At the horizon, the frame-dragging term $\bar{\omega}$ will tend to the constant $\omega_H$ for the reason that DM density vanishes at the horizon.
In the presence of DM, the frame-dragging term will decrease compared with the Kerr case.
We also compare the numerical result of spherically symmetric GBH with the Schwarzchild BH in Fig. \ref{SchvsDMSch}.
In the case of spherical symmetry, the frame-dragging term $\bar{\omega}$ is identically equal to zero.
All the metric elements $(\bar{f},\bar{m},\bar{l},\bar{\omega})$ are not dependent on the angular coordinate $\theta$ and the function $\bar{l}$ is identically equal to the function $\bar{m}$.
Compared with the Schwarzchild metric, the metric value of spherically symmetric GBH is smaller for the functions $(\bar{f},\bar{m})$.
The behavior of metric functions for the rotating case is different from the non-rotating case.
Furthermore, the difference in metric functions with DM and without DM is larger for rotating GBH than nonrotating GBH, which indicates that for the same DM distribution rotating GBH will be easier for us to detect DM.

\section{conclusions}\label{sec5}
In this paper, we numerically calculate the rotating galactic black holes based on a relaxed Newton-Raphson method to solve the discretized field equations with Newton's polynomial finite difference scheme.
The numerical code is validated by considering static and axisymmetric black holes without dark matter, and spherically symmetric black holes with dark matter.
Our numerical results for the rotating GBH without dark matter are consistent with analytical solutions for the Kerr case.
The results for the non-rotating GBH with dark matter from Cardoso's method are also consistent with our numerical results.
Ultimately, we can confidently utilize our numerical code to construct the rotating GBH embedded in dark matter.
It's the first time to extend the spherical galactic black hole to the rotating galactic black hole surrounded by the Einstein cluster.
For the same dark matter distribution, the rotating black hole causes the larger difference between the Kerr black hole and a galactic black hole.
The rotating galactic black hole with dark matter may make it easier for us to detect dark matter distribution.
Our findings provide a solid groundwork for future studies on the shadows, quasi-normal modes, extreme mass ratio inspirals, and other phenomena associated with rotating galactic black holes.
We plan to address these questions in future work.
Our numerical code is based on the open-source code developed by Andrew Sullivan et al at \url{https://github.com/sullivanandrew/XPDES}.
The open-source code of this paper is available at \url{https://github.com/ChaoZhangCode/GBH.git}.

\begin{acknowledgments}
We thank Andrew Sullivan et al for providing the open-source code in GitHub.
This work was supported by the China Postdoctoral Science Foundation under Grant No. 2023M742297,
the Natural Science Foundation of China under Grant No. 12347159, and the China Postdoctoral Science Foundation under Grant No. BX20220313.
\end{acknowledgments}

\appendix
\section{The Kerr metric in isotropic coordinates}\label{KERRICS}
The Kerr metric in isotropic coordinates is
\begin{equation}
\begin{split}
f_{\rm GR} &= \left( 1 - \frac{\rho_{H}^2}{\rho^2}\right)^2 \frac{F_1}{F_2}, \\
m_{\rm GR} &= \left( 1 - \frac{\rho_{H}^2}{\rho^2}\right)^2\frac{F_1^2}{F_2}, \\
l_{\rm GR} &= \left( 1 - \frac{\rho_{H}^2}{\rho^2}\right)^2, \\
\omega_{\rm GR} &= \frac{F_3}{F_2},
\end{split}
\end{equation}
where
\begin{equation}
\begin{split}
\label{eq:F123def}
F_1 &= \frac{2 M_0^2}{\rho^2} +\left( 1 - \frac{\rho_{H}^2}{\rho^2}\right)^2 + \frac{2 M_0}{\rho} \left( 1 + \frac{\rho_{H}^2}{\rho^2}\right) \\
&- \frac{M_0^2 - 4 \rho_{H}^2}{\rho^2} \sin^2\theta, \\
F_2 &= \left[ \frac{2 M_0^2}{\rho^2} +\left( 1 - \frac{\rho_{H}^2}{\rho^2}\right)^2+ \frac{2 M_0}{\rho} \left( 1 +\frac{\rho_{H}^2}{\rho^2}\right) \right]^2 \\
&- \left( 1 - \frac{\rho_{H}^2}{\rho^2}\right)^2 \frac{M_0^2 - 4 \rho_{H}^2}{\rho^2} \sin^2\theta, \\
F_3 &= \frac{2 M_0 \sqrt{M_0^2 - 4 \rho_{H}^2} \left( 1+\frac{M_0}{\rho}+\frac{\rho_{H}^2}{\rho^2}\right)^2}{\rho^2}.
\end{split}
\end{equation}

\section{The spherically symmetric black hole in dark matter}\label{SSCGBH}
We study the static, spherically-symmetric spacetime describing a BH immersed in dark matter halo, with line element
\begin{equation}
ds^2=-a(\rho) dt^2+b(\rho) \left( d \rho^2 + \rho^2 d \theta^2+ \rho^2\sin^2\theta d \phi^2\right),
\end{equation}
and characterized by a stress tensor
\begin{equation}
T^{\mu}_{~\nu}={\rm diag}(-\epsilon,0, p_t,p_t), 
\end{equation}
where $\epsilon$ is the energy density of the fluid, $p_t$ is the tangential pressure.
We can derive two independent equations from the Einstein field equations
\begin{equation}\label{SSC1}
\frac{4 \rho b(\rho) b''(\rho)-3 \rho b'(\rho)^2+8 b(\rho) b'(\rho)}{4 \rho b(\rho)^3}+\epsilon(\rho)=0,
\end{equation}
\begin{equation}\label{SSC2}
\frac{2 b(\rho) \left(\rho a'(\rho)+2 a(\rho)\right) b'(\rho)+4 b(\rho)^2 a'(\rho)+\rho a(\rho) b'(\rho)^2}{4 \rho a(\rho) b(\rho)^3}=0,    
\end{equation}
where the prime denotes the derivative with respect to the radius $\rho$.
We can also make the substitution 
\begin{equation}
    x=1-\frac{\rho_H}{\rho}.
\end{equation}
To obtain asymptotically flat solutions with a regular event horizon and with the proper symmetries, the boundary at the horizon is
\begin{equation}
   a|_{x=0}=0,
\end{equation}
and at the infinity are 
\begin{equation}
 a|_{x=1}=1 ,\qquad b|_{x=1}=1.
\end{equation}
For a given DM density $\epsilon(x)$, we can use Mathematica to solve the Eqs. \eqref{SSC1} and \eqref{SSC2} to get the metric elements $a(x)$ and $b(x)$.

%

\end{document}